\documentclass[12pt]{article}%
\usepackage{amssymb}
\usepackage{amsfonts}
\usepackage{amsmath}
\usepackage{graphicx}%
\providecommand{\U}[1]{\protect\rule{.1in}{.1in}}

\makeatletter
\renewcommand{\section}{\@startsection{section}{1}{0pt}{-3.5ex plus -1ex minus -.2ex}{2.3ex plus.2ex}{\centering\Large\bfseries}} \makeatother
\fontsize{14}{14pt}\selectfont
\begin{document}

\title{\textbf{{\large On the Wave Zone of Synchrotron Radiation}}}
\author{V.G. Bagrov \footnote{bagrov@phys.tsu.ru} \\Tomsk State University, 634050, Tomsk, Lenin Ave. 36, Russia, \\Institute of High Current Electronics, SB RAS \\634055, Tomsk, Akademichesky Ave. 4, Russia}

\maketitle

\begin{abstract}
	The extension of the wave zone of synchrotron radiation is studied. \vspace{2mm}
\end{abstract}

A theoretical study into the problem of the wave zone of synchrotron radiation
was carried out in Ref.~\cite{1} (pp.~45--49). This research was published in
English two years later [2] (pp.~34--37). Since then, for half a century,
these results (concerning the issue of wave radiation zone) have been
considered as rigorously proven. Indeed, the proof given in Refs.~\cite{1,2}
remains valid; however, the possibility of applying these reasonings to an
estimation of the wave zone size has to be refined. This problem is the
subject of the present work.

The radiation power $W$ of a charged particle in classical electrodynamics is
given by the expression%
\begin{equation}
W=\oint_{\mathbf{s}}(\mathbf{S}d\mathbf{s}), \label{a.1}%
\end{equation}
where the surface $\mathbf{s}$ is a sphere with its center at the charge
location, and the vector $\mathbf{S}$ is the Poynting vector%
\begin{equation}
\mathbf{S}=\frac{c}{4\pi}\left[  \mathbf{E}\mathbf{H}\right]  . \label{a.2}%
\end{equation}
The electric $\mathbf{E}$ and magnetic $\mathbf{H}$ radiation fields of a
point charge, determined with the help of Li\'{e}nard--Wiechert potentials,
can be written in the form%
\begin{equation}
\mathbf{E}=\frac{e}{R^{2}\left[  1-(\mathbf{n}\mbox{\boldmath$\beta$})\right]
^{3}}\left\{  (1-\beta^{2})(\mathbf{n}-\mbox{\boldmath$\beta$})+\frac{R}%
{c}\left[  \mathbf{n}\left[  (\mathbf{n}%
-\mbox{\boldmath$\beta$})\mbox{\boldmath$\dot{\beta}$}\right]  \right]
\right\}  ,\nonumber
\end{equation}%
\begin{equation}
\mathbf{H}=\left[  \mathbf{n}\mathbf{E}\right]  ,\ \ \mathbf{n}=\frac
{\mathbf{R}}{R}\,\,. \label{a.3}%
\end{equation}
Here, $\mathbf{v}=c\mbox{\boldmath$\beta$}$ is the charge velocity;
$\mathbf{R}$ is the vector connecting the location point of the radiating
charge to the observation point of radiation; $e$ is the charge magnitude; $c$
is the speed of light. In the right-hand sides of relations~(\ref{a.3}), all
the quantities are regarded at the time instant $\tau$ determined by the
condition%
\begin{equation}
\tau+\frac{R(\tau)}{c}=t\,\,. \label{a.4}%
\end{equation}
In particular, Eq.~(\ref{a.4}) implies%
\begin{equation}
\frac{\partial t}{\partial\tau}=1-(\mathbf{n}\mbox{\boldmath$\beta$})\,\,.
\label{a.5}%
\end{equation}

\begin{figure}
[ptb]
\begin{center}
\includegraphics[
natheight=7.208200in,
natwidth=8.666300in,
height=3.2958in,
width=3.9565in
]%
{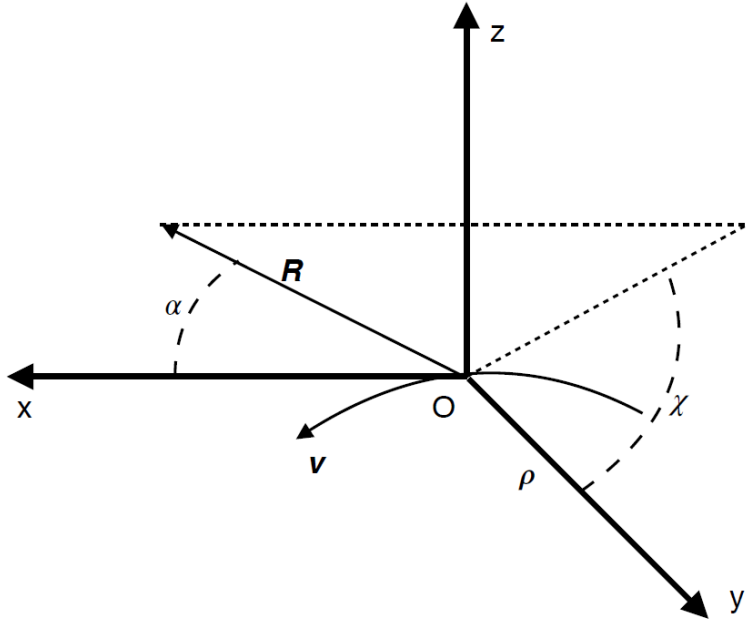}%
\caption{Coordinate system.}%
\label{fig}%
\end{center}
\end{figure}

When considering synchrotron radiation, the velocity and acceleration of a
particle are mutually orthogonal, constant in absolute value, and belong to
one and the same fixed plane. Let us select the following coordinate system
(see Fig.~\ref{fig}). The origin of the coordinate system is chosen at the
location point of the radiating charge. The $x$-axis is directed along the
velocity of the electron, and the $y$-axis is directed toward the center of
the circular trajectory. We choose the $z$-axis so that the coordinate system
is right-handed (when the charge moves in a constant and uniform magnetic
field, with the $z$-axis being parallel to the external magnetic field). The
radius of a circular orbit of the radiating particle is denoted by $\rho$ and
is oriented along the $y$-axis of the given coordinate system. The angle
formed by the $x$-axis and the vector $\mathbf{R}$ is denoted by $\alpha$
$(0\leqslant\alpha\leqslant\pi)$, while the angle between the $y$-axis and the
projection of the vector $\mathbf{R}$ onto the $yz$-plane is denoted by $\chi$
$(0\leqslant\chi<2\pi)$. The surface element is given by $d\mathbf{s}%
=\mathbf{R}Rd\Omega,\,d\Omega=\sin\alpha d\alpha d\chi$.%

Taking into account the above notation and Eqs.~(\ref{a.1}) and (\ref{a.2}),
one easily obtains a well-known expression [1, 2] for the instantaneous power
of synchrotron radiation,%
\begin{equation}
W=\frac{ce^{2}\beta^{4}}{4\pi\rho^{2}}\int_{0}^{\pi}\int_{0}^{2\pi}\psi
\sin\alpha\,d\alpha\,d\chi\,, \label{a.6}%
\end{equation}
where, in virtue of relations (\ref{a.4}) and (\ref{a.5}), the right- and
left-hand sides of Eq.~(\ref{a.6}) are regarded at the radiation instant
$\tau$, with the following notation being used:
\begin{equation}
\psi=a_{0}(\beta;\alpha,\chi)+a_{1}(\beta;\alpha,\chi)k+a_{2}(\beta
;\alpha,\chi)k^{2}\,;\nonumber
\end{equation}%
\begin{equation}
a_{0}=\frac{(\beta-\cos\alpha)^{2}+(1-\beta^{2})\sin^{2}\alpha\sin^{2}\chi
}{(1-\beta\cos\alpha)^{5}}\,,\nonumber
\end{equation}%
\begin{equation}
a_{1}=\frac{2\sin\alpha(\beta-\cos\alpha)\cos\chi}{(1-\beta\cos\alpha)^{5}%
}\,,\ \ a_{2}=\frac{\sin^{2}\alpha}{(1-\beta\cos\alpha)^{5}}\,; \label{a.7}%
\end{equation}%
\begin{equation}
k=\frac{1-\beta^{2}}{\beta}\frac{\rho}{R}\,\,. \label{a.8}%
\end{equation}

The expression in Eq.~(\ref{a.7}) is a second-order polynomial in the
parameter $k$ with three coefficients, $a_{s}(\beta;\alpha,\chi)$ ($s=0,1,2$),
depending on the value of $\beta$ and on the integration angles $\alpha,\chi$.
For $R\rightarrow\infty$ (according to Eq.~(\ref{a.8}), this corresponds to
the limit $k\rightarrow0$), the given expression results in the instantaneous
angular distribution $a_{0}(\beta;\alpha,\chi)$ of synchrotron radiation power.

Formally, the value of $k$ in Eq. (\ref{a.7}) is an expansion parameter, and
therefore in\ Refs.~\cite{1,2} we investigated the region $k<1$, with the
tacit assumption that this is the radiation wave zone. However, the smallness
of the parameter $k$ and the definition of the wave zone are related only
indirectly: the parameter $k$ may be small, but the wave zone has not yet been formed.

Let us point out two obvious facts: a) the specific structure of the
coefficients $a_{s}(\beta;\alpha,\chi)$ may alter the conclusion about the
boundaries of the wave radiation zone; however, this structure was not
investigated in Refs.~\cite{1,2}; b) if we consider radiation in a certain
cone (the cone is given by the region of integration with respect to
$\alpha,\chi$), then the wave zone is determined not only by the radiation
properties, but also by the choice of the cone in question. This fact was not
considered in Refs.~\cite{1,2}, either.

Let us examine the total power of synchrotron radiation. Substituting the
expressions (\ref{a.7}) and (\ref{a.8}) into Eq.~(\ref{a.6}), we carry out
integration over $\alpha,\chi$ in the entire space. As a result of simple
calculations, we find
\begin{equation}
W=W_{0}\Phi_{0},\ \ W_{0}=\frac{2ce^{2}\beta^{4}}{3\rho^{2}(1-\beta^{2})^{2}%
},\ \ \Phi_{0}=1+{\overline{k}}\,^{2}\,,\ \ \overline{k}=\sqrt{\frac
{1-\beta^{2}}{\beta^{2}}}\frac{\rho}{R}\,. \label{a.9}%
\end{equation}
Here, $W_{0}$ is the total emitted power in the limit $R\rightarrow\infty$ [1,
2], and the coefficient $\overline{k}$ characterizes the size of the
near-field zone.

We assume the space region $\overline{k}<1$ to be the radiation wave zone.
From Eqs.~(\ref{a.8}) and (\ref{a.9}), it then follows that $k$ and
$\overline{k}$ are related by
\begin{equation}
\gamma k=\overline{k}\,,\ \ \gamma=\frac{1}{\sqrt{1-\beta^{2}}}\,,
\label{a.10}%
\end{equation}
where $\gamma$ is the relativistic factor (assuming, for instance, the
electron energy to be $\sim$1 GeV, we obtain for the relativistic factor the
approximation $\gamma\sim2000$). Eq.~(\ref{a.10}) implies that $k<\overline
{k}$, and also the fact that $\overline{k}<1$ is not necessarily the case at
$k<1$. Consequently, the condition $k<1$, discussed in Refs.~\cite{1,2}, is
only a necessary one, but is not sufficient to determine the synchrotron
radiation wave zone ($k<\overline{k}<1$).

From Eq. (\ref{a.9}), it follows that%
\begin{equation}
\overline{k}=\frac{R_{v}}{R}\,,\ \ R_{v}=\sqrt{\frac{1-\beta^{2}}{\beta^{2}}%
}\,\rho=\frac{\rho}{\sqrt{\gamma^{2}-1}}\,, \label{a.11}%
\end{equation}
where $R_{v}$ is the distance separating the radiation space into the wave
zone ($R_{v}<R$) and the near-field zone ($R_{v}>R$). The classical theory [1,
2] predicts a relation between the orbit radius $\rho$, the velocity of the
radiating particle, and the strength $H$ of the controlling external magnetic
field:%
\begin{equation}
\rho=\sqrt{\frac{\beta^{2}}{1-\beta^{2}}}\,\frac{m_{0}c^{2}}{|eH|}%
=\sqrt{\gamma^{2}-1}\,\frac{m_{0}c^{2}}{|eH|}\,, \label{a.12}%
\end{equation}
where $m_{0}$ is the rest mass of the radiating particle. Substituting the
expression (\ref{a.12}) into Eq. (\ref{a.11}), we obtain $R_{v}$ in the form
\begin{equation}
R_{v}=\frac{m_{0}c^{2}}{|eH|}=\frac{c}{\omega_{c}}\,,\ \ \omega_{c}%
=\frac{|eH|}{m_{0}c}\,, \label{a.13}%
\end{equation}
where $\omega_{c}$ is cyclotron frequency. For instance, the field
$H=1\,T=10000$ results in $R_{v}=0,1704$ cm.
The expression (\ref{a.13}) shows that the parameter $R_{v}$ is determined
only by the rest energy of the radiating particle and by the magnitude of the
external magnetic field. The other characteristics (for example, speed) of the
radiating particle do not influence the quantity $R_{v}$.

Let us consider two examples when the radiation cone does not cover the entire space.

As a first example, we represent the whole space as the sum of two $(\pm)$
subspaces: the subspace $(-)$ is given by the region $y\geqslant0$ (which
corresponds to $\cos\chi\geqslant0$, and the center of the orbit lies in this
subspace); the subspace $(+)$ is given by the region $y<0$ (which corresponds
to $\cos\chi<0$, and the center of the orbit does not lie in this subspace).
It is easy to obtain the expressions%
\begin{equation}
W_{(\pm)}=W_{0}\Phi_{(\pm)},\ \ \Phi_{(\pm)}=\frac{1}{2}\left(  1+{\overline
{k}}\,^{2}\mp\frac{3\beta}{8}\overline{k}\right)  ,\ \ \Phi_{0}=\Phi
_{(+)}+\Phi_{(-)}\,. \label{a.14}%
\end{equation}
As expected, the particle characteristics (being its speed $\beta$) now also
enter the expressions (\ref{a.14}) for $\Phi_{(\pm)}$; however, the condition
$\overline{k}<1$ determines the wave radiation zone in this case as well.

As a second example, we represent the entire space as the sum of two,
$(\mathrm{in})$ and $(\mathrm{out})$, subspaces: the $(\mathrm{in})$-subspace
corresponds to the radiation value $(\mbox{\boldmath$\beta$}\mathbf{S})>0$
(which determines the range of variation $0<\alpha<\pi/2$); the $(\mathrm{out}%
)$-subspace corresponds to the radiation value
$(\mbox{\boldmath$\beta$}\mathbf{S})<0$ (which determines the range of
variation $\pi/2<\alpha<\pi$). Thus, the space is divided into the
$(\mathrm{in})$-part, in which the projection of radiation pulse on the
particle velocity is positive, and the $(\mathrm{out})$-part, in which the
projection of radiation pulse on the particle velocity is negative.

For the $(\mathrm{in})$-subspace, it is easy to obtain expression%
\begin{equation}
W_{(\mathrm{in})}=W_{0}\Phi_{(\mathrm{in})},\ \ \Phi_{(\mathrm{in})}%
=\Phi_{(\mathrm{in})}^{0}(\beta)\left[  1+A_{(\mathrm{in})}(\beta
)\,{\overline{k}}\,^{2}\right]  ,\nonumber
\end{equation}%
\begin{equation}
\Phi_{(\mathrm{in})}^{0}(\beta)=\frac{16+\beta(3+\beta^{2})(7-3\beta^{2})}%
{32},\ \ A_{(\mathrm{in})}(\beta)=\frac{2(8-9\beta+3\beta^{2})(1+\beta
)}{16-11\beta+6\beta^{2}-3\beta^{3}}\,. \label{a.15}%
\end{equation}
We note the simplest properties of the functions $\Phi_{(\mathrm{in})}%
^{0}(\beta)$ and $A_{(\mathrm{in})}(\beta)$ in the segment $0\leqslant
\beta\leqslant1$. The function $\Phi_{(\mathrm{in})}^{0}(\beta)$ increases
monotonously,
\begin{equation}
\Phi_{(\mathrm{in})}^{0}(0)=\frac{1}{2}\leqslant\Phi_{(\mathrm{in})}^{0}%
(\beta)\leqslant1=\Phi_{(\mathrm{in})}^{0}(1)\,, \label{a.16}%
\end{equation}
and the function $A_{(\mathrm{in})}(\beta)$ at the ends of the segment
$A_{(\mathrm{in})}(0)=A_{(\mathrm{in})}(1)=1$ has a unique maximum at the
point%
\begin{equation}
\beta_{\mathrm{\max}}=\frac{3-\sqrt{5}}{2}\approx0,38197\,,\ \ A_{(\mathrm{in}%
)}(\beta_{\mathrm{\max}})=\frac{10(18\sqrt{5}-5)}{319}\approx1,2617\,.
\label{a.17}%
\end{equation}
Thus, the variation domain of the function $A_{(\mathrm{in})}(\beta)$ is
bounded:%
\begin{equation}
1\leqslant A_{(\mathrm{in})}(\beta)\leqslant\frac{10(18\sqrt{5}-5)}%
{319}\approx1,2617\,. \label{a.18}%
\end{equation}

Returning to the expressions (\ref{a.15}) and taking into account
Eqs.~(\ref{a.16})--(\ref{a.18}), we get%
\begin{equation}
W_{(\mathrm{in})}=\left\{
\begin{array}
[c]{cc}%
\frac{1}{2}\,W_{0}\Phi_{0}\ \ \mbox{non-relativistic limit $(\beta \ll 1)$}; &
\\
W_{0}\Phi_{0}\ \ \mbox{ultra-relativistic limit $(1 \ll \gamma)$}. &
\end{array}
\right.  \label{a.19}%
\end{equation}
From Eq. (\ref{a.19}), we find that in the nonrelativistic case half of the
emtted energy falls on the $(\mathrm{in})$-subspace, whereas in the
ultra-relativistic case this subspace accumulates all the emitted energy. The
emergence condition for the wave zone still has the form $\overline{k}<1$.

For the $(\mathrm{out})$-subspace, we obtain the expression%
\begin{equation}
W_{(\mathrm{out})}=W_{0}\Phi_{(\mathrm{out})},\ \ \Phi_{(\mathrm{out})}%
=\Phi_{(\mathrm{out})}^{0}(\beta)\left[  1+A_{(\mathrm{out})}(\beta
)\,k\,^{2}\right]  ,\nonumber
\end{equation}%
\begin{equation}
\Phi_{(\mathrm{out})}^{0}(\beta)=\frac{16-\beta(3+\beta^{2})(7-3\beta^{2}%
)}{32},\ \ A_{(\mathrm{out})}(\beta)=\frac{2(8+9\beta+3\beta^{2})}%
{(1+\beta)(16+11\beta+6\beta^{2}+3\beta^{3})}\,. \label{a.20}%
\end{equation}
From Eqs. (\ref{a.15}) and (\ref{a.20}), it follows that $\Phi_{0}%
=\Phi_{(\mathrm{in})}+\Phi_{(\mathrm{out})}$, which corresponds to the
equality of the total radiation energy to the sum of the two energies in the
subspaces. This also means $1=\Phi_{(\mathrm{in})}^{0}(\beta)+\Phi
_{(\mathrm{out})}^{0}(\beta)$, which indicates that the radiation energy in
the wave zone is equal to the sum of the two energies in the wave zone of the
subspaces. The functions $\Phi_{(\mathrm{out})}^{0}(\beta)$ and
$A_{(\mathrm{out})}(\beta)$ in the segment $0\leqslant\beta\leqslant1$ are
bounded and monotonously decreasing functions of $\beta$:%
\begin{equation}
\frac{1}{2}\geqslant\Phi_{(\mathrm{out})}^{0}(\beta)\geqslant
0\,,\ \ 1\geqslant A_{(\mathrm{out})}(\beta)\geqslant\frac{5}{9}\,.
\label{a.21}%
\end{equation}

In the nonrelativistic limit $(\beta\ll1)$, as implied by Eqs. (\ref{a.21})
and (\ref{a.19}), we find $W_{(\mathrm{out})}=W_{(\mathrm{in})}=\frac{1}%
{2}W_{0}\Phi_{0}$ (in this case, with allowance made for Eq.~(\ref{a.10}), it
has been taken into account that $k=\overline{k}$).

In the ultra-relativistic limit $(1\ll\gamma)$ for $A_{(\mathrm{out})}%
(\beta\sim1)\approx\frac{5}{9}$, with account taken of Eq. (\ref{a.20}), we
find%
\begin{equation}
\Phi_{(\mathrm{out})}=\Phi_{(\mathrm{out})}^{0}(\beta\sim1)\left(  1+\frac
{5}{9}\,k\,^{2}\right)  , \label{a.22}%
\end{equation}
which differs significantly from the expression (\ref{a.9}) involving $k$. Eq.
(\ref{a.22}) involves $k$, and in this case the wave zone emerges earlier than
in Eq.~(\ref{a.9}). However, as noted above, the limit $\Phi_{(\mathrm{out}%
)}^{0}(\beta\sim1)\rightarrow0$ holds true, and, according to Eq.~(\ref{a.20}%
), this results in $\Phi_{(\mathrm{out})}(\beta\sim1)\rightarrow0$.
Consequently, the radiation value $W_{(\mathrm{out})}$ in the
ultra-relativistic limit is extremely small in relation to $W_{(\mathrm{in})}$
and does not offer any tangible contribution to the total radiation.

For a given external magnetic field $H$ in the ultra-relativistic limit, the
radiation value $W_{(\mathrm{out})}$ is not only small in relation to
$W_{(\mathrm{in})}$, but also tends to zero as the energy of the radiating
charge increases. Indeed, taking into account Eq.~(\ref{a.9}) for the quantity
$W_{0}$ and using Eq.~(\ref{a.12}), we find the expression%
\begin{equation}
W_{0}=\frac{2e^{2}\beta^{2}\omega_{c}^{2}}{3c(1-\beta^{2})}=\frac{2e^{2}%
\omega_{c}^{2}}{3c}(\gamma^{2}-1)\,, \label{a.23}%
\end{equation}
which implies that the radiation value $W_{0}$ (and thereby also the value of
$W_{\left(  \mathrm{in}\right)  }$) for some fixed $H$ (with a fixed
$\omega_{c}$) increases infinitely with an increasing charge energy
($\sim\gamma^{2}$). For $W_{(\mathrm{out})}$ in Eq.~(\ref{a.20}), we get
\begin{equation}
W_{(\mathrm{out})}=\frac{e^{2}\omega_{c}^{2}}{c}\beta^{2}(1-\beta
^{2})B_{(\mathrm{out})}(\beta)\left[  1+A_{(\mathrm{out})}(\beta
)\,k\,^{2}\right]  ,\nonumber
\end{equation}%
\begin{equation}
B_{(\mathrm{out})}(\beta)=\frac{16+11\beta+6\beta^{2}+3\beta^{3}}%
{48(1+\beta)^{2}}\,. \label{a.24}%
\end{equation}
The value of $B_{(\mathrm{out})}(\beta)$, decreasing monotonously in the
segment $0\leqslant\beta\leqslant1$, is bounded: $\frac{1}{3}\geqslant
B_{(\mathrm{out})}(\beta)\geqslant\frac{3}{16}$. From Eq.~(\ref{a.24}), it
follows that a fixed $H$ results in a radiation value $W_{(\mathrm{out})}$
decreasing to zero as the energy $(\sim\gamma^{-2})$\ increases.

It has been shown that the wave zone of synchrotron radiation is determined by
the condition $\overline{k}<1$ (expressions (\ref{a.9}) or (\ref{a.11})),
which differs from the condition of Refs.~\cite{1,2}.

\paragraph{Acknowledgments.}
This work was supported in part by the RFBR (grant No. 18-02-00149) and by the Program for Improving TSU's Competitiveness among World Leading Scientific and Educational Centers.

\end{document}